\documentclass[aps,prb,preprint,superscriptaddress,showpacs,floatfix]{revtex4-1}
\usepackage{graphicx}
\usepackage{amsmath,amssymb}
\begin{document}
\title{Hybrid cluster+RG approach to the theory of
phase transitions in strongly coupled Landau-Ginzburg-Wilson model}
\author{V. I. Tokar}
\affiliation{IPCMS, Universit{\'e} de Strasbourg--CNRS, UMR 7504,
23 rue du Loess, F-67034 Strasbourg, France}
\date{\today}
\begin{abstract}
It is argued that cluster methods provide a viable alternative to Wilson's
momentum shell integration technique at the early stage of renormalization
in the field-theoretic models with strongly coupled fields because these
methods allow for systematic accounting of all interactions in the system
irrespective of their strength.  These methods, however, are restricted
to relatively small spatial scales, so they ought to be supplemented
with more conventional renormalization-group (RG) techniques to account
for large scale correlations.  To fulfill this goal a ``layer-cake''
renormalization scheme earlier developed for rotationally symmetric
Hamiltonians has been generalized to the lattice case.  The RG technique
can be naturally integrated with an appropriately modified cluster
method so that the RG equations were used only in the presence of large
scale fluctuations, while in their absence the approach reduced to a
conventional cluster method.  As an illustrative example the simplest
single-site cluster approximation together with the local-potential RG
equation are applied to simple cubic Ising model to calculate several
non-universal quantities such as the magnetization curve, the critical
temperature, and some critical amplitudes.  Good agreement with Monte
Carlo simulations and series expansion results was found.
\end{abstract}
%Phase transitions in statistical mechanics and thermodynamics, 05.70.Fh
%Phase diagrams metals and alloys, 81.30.Bx
%Renormalization-group theory in phase transitions, 64.60.ae
%Ising model
%   lattice theory, 05.50.+q
%   magnetic ordering, 75.10.Hk
\maketitle 
\section{Introduction}
Modern theory of critical phenomena in field-theoretic
systems is based on the perturbative treatment of the
Landau-Ginzburg-Wilson (LGW) model with weakly interacting fluctuating
fields.\cite{wilson,le_guillou_zinn-justin,3DIsing2,free_params2RG,free_params2RG2} The
theory is adequate in the critical region, at least in the space
dimensions sufficiently close to four, where the interactions become
weak at the late stages of the renormalization irrespective of
whether the initial interactions were weak or strong.\cite{wilson}
In the latter case, however, the numerical values of the majority of
non-universal quantities,---such as the critical temperatures and critical
amplitudes,---remain largely unknown\cite{free_params2RG,free_params2RG2}
because non-perturbative field-theoretic techniques that are necessary
to treat the strong coupling case are much less reliable than the
perturbation theory.

There exist, however, strongly-coupled LGW-type models of considerable
practical importance.  For example, the Ising model that is widely used in
the description of alloy thermodynamics (see Ref.\ \onlinecite{ducastelle}
and references therein) can be cast in the form of strongly coupled LGW
model, as will be discussed in Section \ref{the_model} below.  Therefore,
the quantitative theory of critical phenomena in strongly coupled LGW-type
models could be used, in particular, in prediction of alloy's phase
diagrams on the basis of {\em ab initio} band-structure calculations.
\cite{ECI,ducastelle,gautier,c_INdep_CE2,Zunger1994,PhysRevLett.92.255702}

In terms of the RG theory a major obstacle hampering a fully quantitative
description of critical phenomena in the strong coupling case lies
in the absence of reliable methods of renormalization during the
so-called transient regime that takes place at the early stage
of the renormalization procedure.\cite{wilson}   The aim of the
present paper is to suggest a technique that would combine two
non-perturbative methods, namely, the cluster method of Refs.\
\onlinecite{tokar1983,tokar_new_1997,tan_topologically_2011} and the
``layer-cake'' renormalization scheme in the local potential approximation
(LPA) \cite{1984,local_potential} appropriately modified in such a way
that the RG treatment would be invoked only when the system develops
long-range correlations while away from the criticality a simpler cluster
technique would be sufficient.

The material below is distributed as follows. In the next section the
models, basic equations and notation will be introduced; besides, with
the help of the field-theoretic diffusion RG equations\cite{hori1952}
qualitative picture of the renormalization in the strong-coupling regime
will be discussed. In Section \ref{c+rg} the cluster plus RG (C+RG)
technique will be introduced and the simplest single site cluster
approximation (SSA) presented as an input to the ``layer-cake''
RG approach which is explained in Section \ref{LPA}.  Furthermore,
in Section \ref{LPA} the LPA RG equation will be derived as a natural
complement of the SSA and in Section \ref{observables} the formulas for
some thermodynamic quantities in the SSA+LPA approach will be derived.
In Section \ref{numerics} the LPA equation derived in the previous
section is transformed to the form that facilitates its numerical
solution.   In Section \ref{SCIsing} the approach will be illustrated
by the description in the framework of the SSA+LPA of the ferromagnetic
ordering in the simple cubic (SC) Ising model in zero magnetic field.
In Section \ref{conclusion} brief discussion of the results obtained
and some concluding remarks will be given.
\section{\label{the_model}The models and the diffusion RG equations}
In this paper we consider the standard LGW model and the Ising model
as its concrete realization.  The LGW Hamiltonian for the fluctuating
field $\phi$ is
\begin{equation}
	H[h] = \frac{1}{2}\sum_{i,j}\phi_iG^{-1}_{ij}\phi_j + U[\phi] - \sum_ih_i\phi_i,
	\label{H}
\end{equation}
where $i,j$ are indexes of the lattice sites and $U$ is the functional
describing all interactions that are not written down explicitly
in the equation. It also includes an arbitrary term of the form
$-\frac{1}{2}\sum_{i,j}\phi_ir_{ij}\phi_j$ that also enters  in the
inverse of the pair correlation function (CF) in the first sum in Eq.\
(\ref{H}) but with opposite sign, so that $H$ does not depend on $r$.
In the quasi-momentum representation the pair CF reads
\begin{equation}
	G({\bf k})= \frac{1}{\epsilon({\bf k})+r({\bf k})},
	\label{G}
\end{equation}
where for convenience the Fourier-transformed ``bare'' intersite
interaction will be chosen in such a way that
\begin{equation}
	\epsilon({\bf k})|_{\bf k\to 0 }\sim Kk^2.
	\label{eps20}
\end{equation}
This can always be achieved by adding and subtracting a constant
to/from the two terms in the denominator in Eq.\ (\ref{G});  $K$ in Eq.\
(\ref{eps20}) is some proportionality constant. In the case of the Ising
model with interaction $J$ between nearest-neighbor (NN) spins on the
SC lattice that we are going to use in our numerical
calculations throughout the paper
\begin{equation}
	\epsilon({\bf k})=2K(3-\cos k_x -\cos k_y -\cos k_z), 	
	\label{epsilon}
\end{equation}
where
\begin{equation}
	K=J/k_BT.
	\label{K}
\end{equation}
Finally, the last term in Eq.\ (\ref{H}) describes the interaction
with both real and fictitious (``source'') external fields.  This
slightly differs from the cluster approach of Refs.\
\onlinecite{tokar1983,gamma_exp,tokar_new_1997,tan_topologically_2011}
where the physical field and the source field were treated separately,
though formally they are indistinguishable because they enter into the
formalism as the sum
\begin{equation}
	h_i=h_i^{ext}+h_i^{sorce}.
	\label{h=h+h}
\end{equation}  
The reason for this modification of the formalism is that the present
paper is devoted mainly to the RG approach though the cluster part will
remain practically unchanged.   But the magnetization in the RG part will
be treated differently from the cluster part, namely, via the Legendre
transform for the reasons explained below in section \ref{SSA}.

To simplify notation, the energy unit will be chosen in such a way
that the Boltzmann factor $1/k_BT$ were equal to unity.  In this case
the physical temperature in the system can be controlled by a value of
some Hamiltonian parameter.  In the calculations below we will use the
inverse of the parameter $K$ from Eqs.\ (\ref{eps20}), (\ref{epsilon}),
and (\ref{K}) as our dimensionless temperature.

In this notation the generating functional of the CFs reads
\begin{equation}
	Z[h]=\int\prod_id\phi_ie^{-H}.
	\label{Z}
\end{equation}
The partition function is obtained from $Z[h]$ by setting $h$ to be equal
to the external field $h^{ext}$.  As is seen from Eq.\ (\ref{h=h+h}),
this is equivalent to the conventional route of setting the fictitious
source field to zero.  \cite{gamma_exp,1984,tokar_new_1997}.

Now with the use of the standard transformations
\cite{gamma_exp,1984,tokar_new_1997,vasiliev1998,tan_topologically_2011}
the generating functional can be cast in the form
\begin{equation}
	Z[h]=\left(\det G\right)^{1/2}\exp\left(\frac{1}{2}hGh\right)R[Gh],
	\label{ZR}
\end{equation}
where the generating functional of the S-matrix is\cite{hori1952}
\begin{equation}
	R[\phi]=\exp\left(\frac{1}{2}\frac{\partial}{\partial\phi}
	G\frac{\partial}{\partial\phi}\right)e^{-U[\phi]}
	\label{RU}
\end{equation}
For simplicity, in Eqs.\ (\ref{ZR}) and (\ref{RU}) and below we omit
summations over the site indexes by adopting vector-matrix notation in
the $N$-dimensional space  ($N$ is the number of lattice sites) by treating
fields as vectors and pair CFs as matrices.

As is easily seen from Eqs.\ (\ref{ZR}) and (\ref{RU}), from the
assumption that $G$ is the exact correlation function of the system it
follows that \cite{tokar1983,gamma_exp,tokar_new_1997}
\begin{equation}
	\left.\frac{\delta^2 \ln R[\phi]}{\delta\phi_i\delta\phi_j}\right|_{\phi=Gh^{ext}}=0.
	\label{d2rdphi2}
\end{equation} 
Similar to the cluster approach of Refs.\
\onlinecite{tokar1983,gamma_exp,tokar_new_1997,tan_topologically_2011},
this equation will constitute the self-consistency condition for
the calculation of $r({\bf k})$.  In the present paper we restrict
ourselves to the smallest one-site ``clusters'' and to the SSA in the
cluster part of the C+RG approach;  the RG part will be treated in the
LPA---the approximation with quasimomentum-independent vertexes, so $r$
in our explicit calculations will be just some non-negative number.
\subsection{Diffusion RG equations}
In the functional formalism \cite{hori1952,wilson,vasiliev1998,1984}
the renormalization (semi)group can be introduced by representing $G$
in Eq.\ (\ref{RU}) as an integral over some evolution parameter $t$:
\begin{equation}
	G=\int_{t_0}^{t_1}dt G_t(t).
	\label{intG}
\end{equation}
(Here and below the subscripts corresponding to continuous variables will
denote partial derivatives with respect to these variables.)  With the
use of the representation Eq.\ (\ref{intG}) it is easy to see that the
functional Eq.\ (\ref{RU}) can be found as the solution of a
Cauchy problem for the evolution equation for the group generator
that in our case takes the form of the diffusion equation \cite{hori1952}
\begin{equation}
	R_t[\phi,t]=\frac{1}{2}\frac{\delta}{\delta\phi} 
	G_t\frac{\delta}{\delta\phi}R[\phi,t]
	\label{diff_eq}
\end{equation}
with the initial value
\begin{equation}
R[\phi,t_0]=\exp(-U[\phi]).
	\label{R0}
\end{equation}
In principle, any parametrization satisfying Eq.\ (\ref{intG}) can be
used Eq.\ (\ref{diff_eq}). For example, in Ref.\ \onlinecite{hori1952} a
straightforward parametrization was proposed consisting (in our notation)
in multiplication of $G$ by $0\leq t\leq 1$
\begin{equation}
	G({\bf k},t)=G({\bf k})t
	\label{hori_parametrization}
\end{equation}
so that $G_t=G$.  This time-independence of the ``diffusion constant''
allows one to represent the solution of the diffusion equation Eq.\
(\ref{diff_eq}) as the convolution of the initial field distribution Eq.\
(\ref{R0}) with the fundamental solution
\begin{equation}
P[{\bf \phi-\phi^\prime},t]=\prod_{\bf k}\frac{1}{[4\pi tG_t({\bf k}) ]^{N/2}}
\exp\left(-\frac{(\phi-\phi^\prime)_{\bf-k}(\phi-\phi^\prime)_{\bf k} }{4tG_t({\bf k})}\right).
\label{N_dim_eq}
\end{equation}
For our purposes it is important to note that in the vicinity of the
critical point $G_t=G({\bf k})$ at small $|{\bf k}|$ is positive and
large. As we know from the properties of the conventional diffusion,
even the delta-function-like initial probability density distributions
that in our case would describe infinitely strong interactions in Eq.\
(\ref{R0}) can be strongly smeared, given sufficient time and sufficiently
large diffusion constant. In our case this would mean diminishing of
the interaction in the renormalized effective interaction $U$ in Eq.\
(\ref{R0}) to small values.  It is to be stressed, however, that the
smearing strongly depends on both the time of the diffusion and on the
value of the diffusion constant, as can be qualitatively seen from the
fundamental solution Eq.\ (\ref{N_dim_eq}).

Thus, in order the interactions at a given ${\bf k}$ were small, the
effective time--diffusion constant product $tG_t({\bf k})$ in Eq.\
(\ref{intG}) should be large.  The success of the RG approach to
critical phenomena is essentially based on the fact that this condition
is satisfied in the vicinity of the critical point, at least in the
systems of dimension close to four.\cite{wilson}  For example, in Eq.\
(\ref{hori_parametrization}) time varies only from 0 to one. But $G_t$
along the path leading to the singularity at ${\bf k=0}$ remains very
large, in fact, infinite at the criticality.  Thus, the  product $tG_t$
can be very large.  However, away from ${\bf k=0}$ and/or away from the
criticality the interactions in $U$ in Eq.\ (\ref{R0}) may turn out
to be too large to be smeared by the kernel Eq.\ (\ref{N_dim_eq}) to
sufficiently broad probability distribution. In this case renormalized
interactions will remain strong and non-perturbative techniques are
necessary to deal with them.

One may wonder whether the simple parametrization Eq.\
(\ref{hori_parametrization}) with all quasimomenta being treated equally
is adequate for discussion of renormalization in the critical region
where the long-wavelength fluctuations are the most important ones.
In Wilson's original approach the renormalization process proceeds
via a different route when the short-wavelength fluctuation are being
integrated out completely with the long-wavelength field components left
intact.\cite{wilson}   To see that all renormalization trajectories
defined by the parametrization in Eq. (\ref{intG}) are physically
equivalent, let us consider a simple case of a smooth cutoff when
large-quasimomenta components are integrated out faster than the
long wave-length ones, as is done in many renormalization schemes
\cite{bogoli︠u︡bov1959,wilson,polchinski_renormalization_1984}.  For
example, the Pauli–Villars regularization \cite{bogoli︠u︡bov1959}
can be obtained within our approach with the use of parametrization
\begin{equation}
G_t({\bf k},t)=\frac{1}{[\epsilon({\bf k})+r({\bf k})+t]^2}.
\label{PV0}
\end{equation}
Integrating this from $t=0$ to some sufficiently large value of $t$
one gets
\begin{equation}
G({\bf k},t)=\frac{1}{\epsilon({\bf k})+r({\bf k})}-\frac{1}{\epsilon({\bf k})+r({\bf k})+t}
\label{PV1}
\end{equation}
In the critical region and at small ${\bf k}$ the denominator in
the first term on the right hand side of  Eq.\ (\ref{G}) is small at
small ${\bf k}$, so in this case $G({\bf k},t)\approx G({\bf k})$.
However, as $\epsilon({\bf k})+r({\bf k})$ grows towards and past the
value of $t$, $G({\bf k},t)$ becomes small thus effectively cutting off
short-wavelength fluctuations in $G({\bf k})$ qualitatively similarly
to the Wilson's method.  From the integrated Eq.\ (\ref{PV0}) one sees
that in case of finite cutoff in $k$-space large values of $t$ larger
than some finite value $t_1$ give contribution into Eq. (\ref{intG})
of $O(1/t_1)$ so for qualitative reasoning they can be neglected. The
interval $[0,t_1]$ can be mapped onto interval [0,1], so qualitatively
the problem reduces to the previously considered case.

This qualitative reasoning was meant to qualitatively substantiate the
observation that the structure of strongly coupled LGW models is such that
during the renormalization flow the system either enters into critical
regime with weak coupling or retains the strong interactions but with
only short-range correlations.  This dichotomy should justify the need
for the hybrid approach when strong coupling short-range behavior is
treated within cluster approximations that are known to be well suited
to this kind of problems while the critical behavior is dealt with
in an RG technique that is the only viable approach to the problem in
field-theoretic models.

Of course, not all LGW-type statistical models are strongly coupled but
some practically important ones may have effective interactions even of
infinite strength.  For example, the Ising model can be transformed into
a field-theoretic system as follows
\begin{eqnarray}
&&Z[h]=\mbox{Tr}_{s_i=\pm1}\exp\left(\frac{1}{2}\sum_{ij}s_iJ_{ij}s_j
+\sum_ih_is_i\right)\nonumber\\
&&=\int\prod_i[d\phi_i2\delta(\phi_i^2-1)]\exp\left(\frac{1}{2}\sum_{ij}\phi_iJ_{ij}\phi_j
+\sum_ih_i\phi_i \right),
\label{ising}
\end{eqnarray}
where the interaction can be represented as infinitely strong
$\lambda\phi^4$ coupling if the delta-functions in Eq.\ (\ref{ising})
are represented as
\begin{equation}
\delta(\phi_i^2-1)=\left.C_{\lambda}e^{-\lambda\sum_i(\phi_i^2-1)^2}
\right|_{\lambda\to\infty},
\label{delta}
\end{equation} ($C_{\lambda}$ is a normalization constant).

The Ising model is being widely used, {\em
inter alia}, in the study of alloy thermodynamics.
\cite{ECI,ducastelle,gautier,c_INdep_CE2,Zunger1994,PhysRevLett.92.255702}
Therefore, the development of efficient techniques of its solution
with realistic interatomic/interspin interactions is a problem of
considerable practical interest.  This problem has been extensively
studied in the past decades and sophisticated techniques has been
developed that predict non-critical behaviors with high accuracy (see
Refs.\ \onlinecite{ducastelle,tokar_new_1997,tan_topologically_2011} and
references therein).  The critical behavior, however, was investigated
only in simple model systems \cite{RG2002review} while in the theory of
alloys the second order phase transitions are usually treated either
within the cluster variation method to which the mean-field critical
behavior is inherent (see Ref.\ \onlinecite{ducastelle} for extensive
bibliography) or with the use of the Monte Carlo (MC) simulations.
\section{\label{c+rg}C+RG method}
\begin{figure}
\begin{center}
\includegraphics[viewport = 0 0 300 220, scale=0.75]{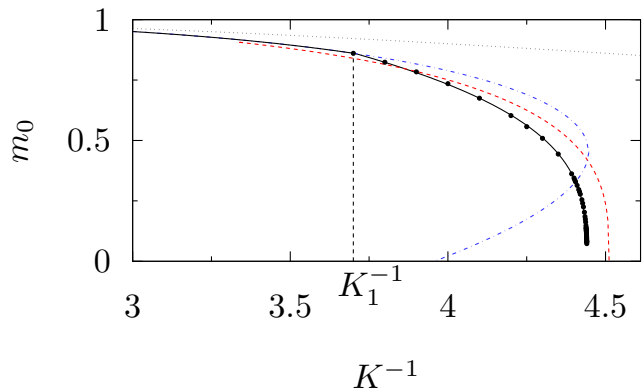}
\end{center}
\caption{\label{fig1} (Color online) Solid line: $K^{-1}$
below $K^{-1}_1$---solution of the SSA Eqs.\ (\ref{SSAm}) and
(\ref{SSAr}); $K^{-1}$ above $K^{-1}_1$---least square fitted cubic
spline\cite{spline_fit} to the black dots obtained from the solutions
of the LPA RG Eq.\ (\ref{the_eq2}); $K^{-1}_1$ is the dimensionless
temperature where the two solutions matched. Dashed curve: the MC
simulation data from Ref.\ \onlinecite{talapov_M(t)}.  Dash-dotted line:
solution of the SSA equations. Dotted curve: Two leading terms of the
low temperature expansion from Ref.\ \onlinecite{lowT_sc}.} \end{figure}

In the functional cluster technique of Refs.\
\onlinecite{tokar1983,gamma_exp,tokar_new_1997,tan_topologically_2011}
one can calculate with good accuracy only the critical temperature
while the critical behavior in general is described very poorly (see,
e.\ g., Fig.\ \ref{fig1}).  This approach, however, is based on the
field-theoretical formalism so in principle should be compatible with
Wilson's RG techniques.  The rest of the present paper will be devoted to
the realization of this observation in application to the strongly coupled
LGW model with the use of the SC Ising model in concrete calculations.

The cluster approach we will use is a systematic technique to
treat lattice field-theoretic models and is well documented in
previous publications, so here it will only be mentioned why
it is insufficient to the description of the second order phase
transitions.  The essence of the approach is easier to explain using
the coarse-grained version \cite{tan_topologically_2011} of the
cluster method though non-coarse-grained version may be more accurate
\cite{tokar1983,tokar_new_1997}.

\begin{figure}
\begin{center}
\includegraphics[viewport = 0 0 300 220, scale=0.75]{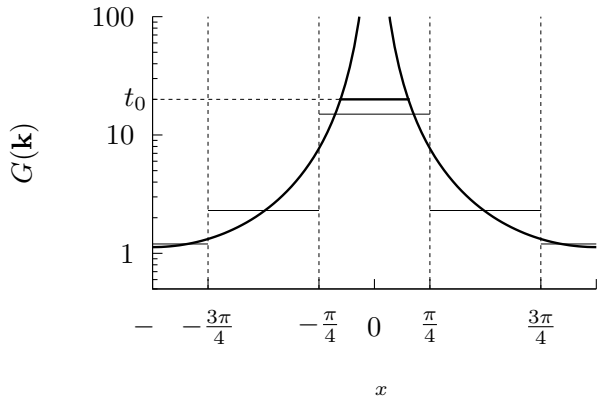}
\end{center}
\caption{\label{fig2}Schematic picture of the separation of $G$ in Eqs.\
(\ref{bar+hat}) and (\ref{GbarGhat}) into $\bar{G}$---thick solid lines
below $t_0$ plus solid ${G}=t_0$ line segment and $\bar{G}$---solid
lines above the ${G}=t_0$ line. Shown also are the average values
of the coarse-grained correlation function (horizontal thin solid
lines).\cite{maier_quantum_2005,tan_topologically_2011} Vertical dashed
lines are the boundaries that divide the initial Brillouin zone into
smaller parts when the system is divided into clusters having four cells
in the $x$-direction.} \end{figure}

In the cluster approach one tries to approximate an infinite system of
strongly coupled fields with a finite cluster of size $N_c$.  The reason
is that the calculation of the partition function Eq.\ (\ref{Z}) can
be approximately reduced to $N_c$-dimensional integral that can be
calculated exactly irrespective of the strength of the interactions.
In Fig.\ \ref{fig2} is drawn a cross-section of the CF $G({\bf k})$ in the
first Brillouin Zone (BZ) of a $d$-dimensional cubic lattice along the
$k_x$ axis.  Let us imagine that the system is divided into clusters of
size $N_c=4^d$ sites.  The BZ of such a system will shrink correspondingly
(see vertical lines). But because the system remains the same we will
keep the initial BZ divided into $4^d$ parts to simplify the discussion.
In the coarse-grained approximation $G({\bf k})$ is approximated by
its average value within each sub-cell (the horizontal lines in Fig.\
\ref{fig2}) and thus $N$ ${\bf k}$-points in the BZ are replaced
by a finite number $N_c$ of the points (for the details see Refs.\
\cite{tan_topologically_2011,maier_quantum_2005}).  The accuracy of the
cluster approximation can be estimated by the size of the deviation of the
exact $G({\bf k})$ from its average values within the subcells. It is
expected that with the growth of $N_c$ the cluster approximation will
be converging to the exact solution as long as the Riemann sum over
subcells will be converging towards the exact integral.

The Riemann sum, however, does not converge in the case of improper
integrals,\cite{riemann} in particular, when the integrand diverges,
as is the case with $G({\bf k})$ at the critical point when $G({\bf
k\to0})\to\infty$.  In this case for an arbitrarily small integration
cell surrounding the point ${\bf k}=0$, the deviation of the exact
$G({\bf k})$ from the average CF will be arbitrarily large at a point
${\bf k}$ sufficiently close to zero.

This necessitates introduction in the vicinity of the critical point of
some other integration technique.  To this end, let us first separate the
singular part of $G({\bf k})$ near ${\bf k\approx0}$ from the nonsingular
behavior at other values of ${\bf k}$ as
\begin{equation}
G({\bf k})=\bar{G}({\bf k})+\hat{G}({\bf k}),
\label{bar+hat}
\end{equation}
where 
\begin{equation}
	\begin{array}{l|c|c}
		&{G}({\bf k})\leq t_0 &{G}({\bf k})>t_0 \\\hline
		\bar{G}({\bf k})= & {G}({\bf k}) & t_0\\\hline
		\hat{G}({\bf k})= & 0 & {G}({\bf k})-t_0
	\end{array}
	\label{GbarGhat}
\end{equation}
(see Fig.\ \ref{fig2}).  It is easy to see that by construction the
variation of $\bar{G}({\bf k})$ is always bounded so the Riemann
sum converge and the problem can be efficiently treated within the
cluster approach.  It is to be noted that far from criticality $r$
in Eq.\ (\ref{G}) can be is large and $G({\bf k})<t_0$.  In this case
the C+RG approach reduces to purely cluster approach of Refs.\
\onlinecite{tokar1983,tokar_new_1997,tan_topologically_2011}.  But when
the second term in Eq.\ (\ref{bar+hat}) is non-zero, the cluster solution
should be augmented with the RG part. The latter will be considered in
the next section while in the rest of this section we will illustrate
the cluster part of the technique with the simple example of the Ising
model in zero external field in the SSA.
\subsection{\label{SSA}SSA in application to the Ising model}
In the SSA the average over ${\bf k}$ is performed over the whole Brillouin zone
and therefore the result is equal to the site-diagonal element of the truncated 
correlation function:
\begin{equation}
\frac{1}{N}\sum_{\bf k}\bar{G}({\bf k})=\bar{G}_{ii}
\label{G_ii}
\end{equation} 
In the present paper we will be interested mainly in the second order
phase transition from the paramagnetic to ferromagnetic state in the
absence of external magnetic fields.  The conventional SSA
\cite{tokar1983,gamma_exp,tokar_new_1997,tan_topologically_2011}
consists in approximating matrix $G$ in Eq.\ (\ref{RU}) by its
diagonal elements in which case the calculation of $R$ reduces to
the calculation of 1D integrals.  But before that one shifts the
integration variables in Eq.\ (\ref{Z}) by the (unknown) value of the
spontaneous magnetization $m$, so the effective potential in the SSA
reads \cite{gamma_exp,tokar_new_1997,tan_topologically_2011}
\begin{equation}
	U^{SSA}[\phi,m]=\sum_i\left(\bar{G}_{ii}^{-1}\phi_i^2/2+m\bar{G}_{ii}^{-1}\phi_i
	-\ln\cosh[\bar{G}_{ii}^{-1}\phi_i+(\bar{G}_{ii}^{-1}-r)m]\right)
	+NC,
	\label{U0}
\end{equation} 
where $C$ does not depend on field $\phi$ and so its explicit value
will not be needed in our present calculations. In case of necessity
it can be easily derived or obtained from the formulas of Ref.\
\onlinecite{tokar_new_1997}.

In the ordered phase the SSA Eq.\ (\ref{U0}) is good at low temperatures
where the two degenerate ground states with magnetizations $\pm m_0$
are energetically well separated, so the state represents a background
of spins directed in one directions plus rare and small clusters of
oppositely directed spins.  

In the critical region, however, the picture is different.  Large clusters
of spins of both directions are almost equal in size with only slight
bias toward one direction.\cite{wilson}  In this case the description of
the system with the use of the average magnetization $m_0\approx0$ does
not represents well the physics of the spontaneous symmetry breaking.
Therefore, in our RG approach we will use Eq.\ (\ref{U0}) with $m=0$
both in the symmetric and the ordered phase and calculate the spontaneous
magnetization via the Legendre transform.\cite{zia,vasiliev1998}

To be more specific, let us assume for the moment that the cut-off
$t_0$ is infinitely large so the RG part is not necessary.  In this case the
problem is solved in the SSA as follows.\cite{gamma_exp,tokar_new_1997}
First, equating to zero the first derivative of Eq.\ (\ref{U0}) with
respect to $\phi_i$ the SSA equation for the equilibrium magnetization
$m_0$ is obtained:
\begin{equation}
m_0=\tanh\left[\left( G^{-1}_{ii}-r\right)m_0\right]
	\label{SSAm}
\end{equation} then Eq.\ (\ref{d2rdphi2}) for $\phi=0$ gives
\begin{equation}
	G_{ii}+m_0^2=1
	\label{SSAr}
\end{equation}
and the SSA magnetization is shown in Fig.\ \ref{fig1}.  As can be seen,
when close to the saturation value $m=1$, the SSA agrees well with the
low-temperature expansion.  It can be shown analytically that Eqs.\
(\ref{SSAm})--(\ref{SSAr}) reproduce exactly its two leading terms.

From Fig.\ \ref{fig1} one can see that close to the critical point,
however, the SSA solution exhibits behavior that is very far from the
correct one as seen, e.\ g., in the MC simulations.\cite{talapov_M(t)}
Therefore, in this region the cluster approximation breaks down and
should be replaced with more adequate treatment.

To apply the C+RG approach one has first to decide on the value of the
cut-off $t_0$ to be used in the calculations.  Assuming that SSA gives
reasonably accurate solution in the non-critical region, let us use
it in assessing the Ginzburg criterion by comparing the square of the
long-wavelength fluctuations of the order parameter with the mean order
parameter squared as\cite{ginzburg_criterion}
\begin{equation}
	\mbox{Gi}\sim \frac{(\Delta m)^2}{m_0^2}=\frac{G({\bf k\to0})}{m_0^2}=\frac{1}{rm_0^2}.	
	\label{Gi}
\end{equation}
Assuming further that the critical region starts at the point where
$\mbox{Gi}\approx1$, one finds with the use of the SSA solution
that at this point (marked by the subscript 0) $1/K_0\approx3.9$ and
$t_0=1/r_0=0.65$---the value we will use in calculations throughout
the paper.  In section \ref{SCIsing} we will show that the results
obtained weakly depend on the choice of the parameter.
\section{\label{LPA}Layer cake renormalization scheme and the LPA} Thus,
having assumed that in the critical region the SSA can satisfactorily
describe only those fluctuations whose correlation do not exceed some
value $t_0$, we also have to develop an RG approach that would take into
account those fluctuations.  Formally we need to calculate an expression
similar to Eq.\ (\ref{RU}) but this time with the effective potential
Eq.\ (\ref{U0}) renormalized by the SSA.  To this end it is convenient
to rewrite the diffusion RG Eq.\ (\ref{diff_eq}) through the logarithm
of the S-matrix\cite{hori_approach_1962} or the effective potential
$U[\phi,t] = -\ln R[\phi,t]$ which we assume can be expanded in the
functional Taylor series in the field $\phi$ as
\begin{equation}\label{U(k)}
	U = N\sum_{n,\{\mathbf{k}_j\}}N^{-n/2}u_n(\{\mathbf{k}_j\},t)
	\delta\left(\sum_{j=1}^n\mathbf{k}_j\right)
\prod_{j=1}^n\phi_{\mathbf{k}_j}.
\end{equation}
In terms of $U$ Eq.\ (\ref{diff_eq}) reads
\begin{equation}
	U_t[\phi,t]= \frac{1}{2}\sum_{\bf k}G_t({\bf k},t)\left(
	\frac{\partial^2 U}{\partial  \phi_{-\bf k}
	\partial \phi_{\bf k}}
	-\frac{\partial U}{\partial  \phi_{-\bf k}}
	\frac{\partial U}{\partial \phi_{\bf k}}
	\right). 
	\label{THE_eq}
\end{equation}
In the layer cake renormalization scheme\cite{1984} the CF $G$ is
represented as the sum of infinitesimal layers of thickness $dt$ which
when stacked together will form the function under consideration (see,
e.\ g., Fig.\ 1 in Ref.\ \onlinecite{1984}).  It is easy to see that
this leads to
\begin{equation}
	G_t({\bf k},t)=\theta\left[G({\bf k},t)-t\right],
	\label{G_t=theta}
\end{equation} 
where $\theta$-function plays the role of the indicator
function\cite{layer_cake} (see Appendix \ref{Layer_Cake}). On substituting
this in Eq.\ (\ref{THE_eq}) together with the expansion Eq.\ (\ref{U(k)})
one should note a subtle difference between the two summations over
$\mathbf{k}$ present in the equation.  The second summation consists
of terms proportional to the product of two lattice delta-functions
(Kronecker symbols)
\begin{equation}
	\sum_{\bf k}\theta\left[ G({\bf k},t)-t\right] 
	\delta\left(\sum_{j=1}^{n-1}\mathbf{k}_j -\mathbf{k}\right)
\delta\left(\mathbf{k}+\sum_{j^\prime=1}^{n^\prime-1}\mathbf{k^\prime}_{j^\prime}\right)
u_n(\{\mathbf{k}_j\}_i,\mathbf{-k},t)
u_{n^\prime}(\{\mathbf{k}^\prime_{j^\prime}\}_{i^\prime},\mathbf{k},t)
	\label{quadratic_term}
\end{equation}
and thus can be performed formally exactly in each term by simply
substituting $\mathbf{k}$ by, say, $\sum_{j=1}^{n-1}\mathbf{k}_j$ in
all factors entering the product. (The subscripts $i,i^\prime$ in Eq.\
(\ref{quadratic_term}) means that corresponding quasimomentum in the set
is absent because it is set to be equal to $\pm\mathbf{k}$; these same
terms reduce by unity the summations over $j,j^\prime$ in the arguments
of the delta-functions.) The first summation in Eq.\ (\ref{THE_eq}),
however, contains only one delta-function
\begin{equation}
	\frac{1}{N}\sum_{\bf k}\theta\left[ G({\bf k},t)-t\right] 
\delta\left(\sum_{j=1}^{n-2}\mathbf{k}_j+\!\not{\mathbf{k}}-\!\not{\mathbf{k}}\right)
u_n(\{\mathbf{k}_j\}_{ii^\prime},\mathbf{k},\mathbf{-k},t)
\label{linear_term}
\end{equation}
so that to carry out the summation one needs to know explicitly the
dependences of the vertex functions on quasimomenta.

Thus, Eq.\ (\ref{THE_eq}) is a complicated nonlinear integro-differential
equation with infinite number of variables and its solution presents a
formidable task. It can be considerably simplified, however, in the LPA
that consists in first setting to zero all $\phi_{\bf k}$ with ${\bf
k}\not=0$ outside the contour defined by the level set $G({\bf k})=t$.
This is legitimate because in our problem only the homogeneous fields
with ${\bf k}=0$ survives after the substitution $\phi=Gh^{ext}$ in Eq.\
(\ref{d2rdphi2}).  After that we assume that the vertex functions $u_n$
are independent of the quasimomenta, so that the functional Eq.\
(\ref{U(k)}) can be mapped onto the function \cite{1984}
\begin{equation}
u(x,t)=\sum_nu_n(t)x^n.
\label{U2u}
\end{equation} 
The approximations should be good in the critical region where small
quasimomenta dominate so that Eq.\ (\ref{U2u}) approximates Eq.\
(\ref{U(k)}) as
\begin{equation}
u_n(t)\approx u_n(\{\mathbf{k}_j\},t)|_{\{\mathbf{k}_j\}\to0}
\label{u2u}
\end{equation}and
\begin{equation}
x\approx \phi_{\mathbf{k}\to0}
\label{x2phi}
\end{equation}
With these approximations it is easy to see that $\theta[G({\bf k\to0},t)]=1$ 
in Eq.\ (\ref{quadratic_term}) and the sum in Eq.\ (\ref{linear_term}) can be calculated
exactly to give
\begin{equation}
	u_t=\frac{1}{2}
	\left[s(t,r)u_{xx}-u_x^2\right],
	\label{the_eq}
\end{equation}
where
\begin{equation}
s(t,r)=n_{SC}(t^{-1}-r)
\label{s}
\end{equation}and $n_{SC}(E)$ is the integrated density of states 
with the dispersion law given by Eq.\ (\ref{epsilon}) (see Appendix
\ref{Layer_Cake}). 

In Fig.\ \ref{fig3} the physical meaning of the function $s$ from Eq.\
(\ref{the_eq}) is explained. One can see, in particular, how the present
approach differs from the usually studied LPA for RG equations in the
scaling form.  \cite{1984,local_potential,caillol_non-perturbative_2012}
The latter is obtained by substituting into Eq.\ (\ref{sum2D(E)})
$D(E)\propto\sqrt{E}$ which is the universal behavior for 3D systems
(see the square root curve in Fig.\ \ref{fig3}).  Fig.\ \ref{fig3}
also illustrates why the conventional perturbative RG is insufficient
for the calculation of non-universal quantities.  While higher powers
of $k_{x,y,z}$ can be neglected in the conventional RG approach due
to their smallness,\cite{wilson} in our case function $s$ contains the
van Hove singularities that appear only when all powers of $k_{x,y,z}$
are kept in the expression for $s$ because the NN interaction in Eq.\
(\ref{epsilon}) that are analytic in ${\bf k}$ have to be summed to
infinite orders to exhibit the singular behavior.
\begin{figure} 
	\begin{center} \includegraphics[viewport = 50 20 300
165, scale=0.75]{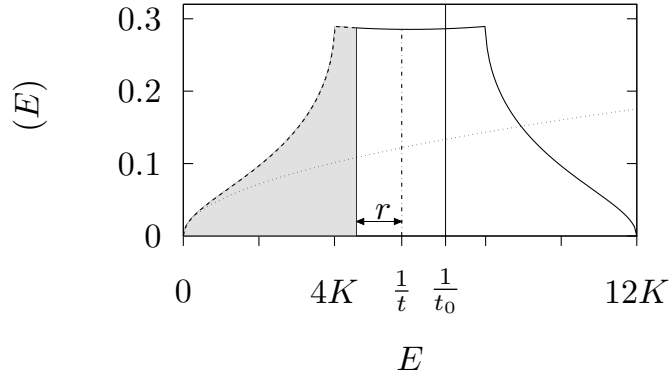} 
        \end{center} 
	\caption{\label{fig3}Density of
states for the SC lattice.  The area of the shaded region is equal to
$s(t)$ in Eqs.\ (\ref{the_eq}) and (\ref{the_eq2}).  The dotted curve
corresponds to the universal square-root scaling behavior observed at
small $E$ and extended on all region of allowed values of $E$.}
\end{figure} 

Thus, by means of the function $s$ the LPA Eq.\ (\ref{the_eq})
is capable of accounting for all pair interactions that are
usually considered to be the most important ones in alloys.
\cite{ECI,ducastelle,gautier,c_INdep_CE2,Zunger1994,EPIs,PhysRevLett.92.255702}
But the aim of the C+RG is to account for all cluster interactions in
the system.  The cluster part of the method is well suited for this
purpose but the RG part needs some adjustment.  It is to be noted that
only quasimomentum-dependent part of cluster interactions poses problems
because the constant part is accounted for in the $u_n(\{\bf k_j\to0\})$
local potential.

A straightforward approach would be to augment the LPA equation Eq.\
(\ref{the_eq}) with RG equations for the coefficients of expansion of
the vertex functions in powers of the quasimomenta.\cite{wilson} The
ensuing equations, however, will be rather cumbersome.  \cite{wilson} So,
presumably, it would be easier to push the initial quasimomentum cutoff
$k_0\sim1/\sqrt{t_0}$ to as small value as possible. This would justify
the neglect of the quasimomentum dependence of most of the vertexes with
the possible exception of the corrections into the pair ones which are
responsible for index $\eta$.  There are two ways to diminish $k_0$.
First,  to enlarge the cluster in the first part of the calculation to
make the BZ possibly smaller and then to chose maximally large $t_0$.
Large $t_0$ will diminish the error due to the neglect of the ${\bf
k}$-dependence of the vertex functions but will enlarge the error in
the cluster part because of larger deviation of $G({\bf k})$ from its
mean value in the cell.  Thus, some optimum choice should exist which
will depend on the presence and the size of non-pair effective cluster
interactions in the system.  In Section \ref{SCIsing} we will show that
the choice of $t_0$ only weakly influences the results at the SSA+LPA
level, so it can be used to improve more advanced calculations with
larger clusters and/or better RG equations.  As can be seen from Eq.\
(\ref{U0}) at $m=0$, the interaction terms are even powers of $\phi_i/\bar{G}_{ii}$
\begin{eqnarray}
	(\phi_i/\bar{G}_{ii})^4/12-(\phi_i/\bar{G}_{ii})^6/45+\dots,
	\label{Un}
\end{eqnarray}so the smaller is the cutoff $t_0$, the smaller is $\bar{G}_{ii}$
hence, larger the interaction.  For example, the coefficient strength
of the $\lambda\phi^4$ interaction at the critical point $1/K_c=4.44$
and cutoff $t_0=0.65$ is
\begin{equation}
	\lambda=[12\bar{G}_{ii}^4(2K_c)^2]^{-1}\approx 3.2
	\label{lambda}
\end{equation}where $\bar{G}_{ii}|_{t_0=0.65}\approx0.6$ and the
$2K_c$ comes from the renormalization of the field $\phi$ necessary to
bring our quadratic term $Kk^2|\phi_{\bf k}|^2$ in our Hamiltonian
(see Eqs.\ (\ref{H}) and (\ref{eps20})) to the canonical form
$k^2|\phi_{\bf k}|^2/2$\cite{wilson,le_guillou_zinn-justin}.  As we
see, the coupling in the case $t_0=0.65$ is rather strong.  This does
not matter in the LPA that we use in the present paper because the
RG equation was derived without any resort to the smallness of the
couplings and thus can be applied to the strong coupling case of
Eq.\ (\ref{lambda}).  However, the LPA is not easy to improve in the
nonperturbative manner to exhibit, for example, more accurate critical
behavior with improved indexes.\cite{local_potential} Therefore, it
would be more practical to push the cutoff $t_0$ higher in which case
$\bar{G}_{ii}$ will approach unity (see Eq.\ (\ref{SSAr}) with $m_0=0$)
and the coupling in Eq.\ (\ref{lambda}) will tend to much smaller value
$\lambda\approx0.4$.  This should facilitate application of perturbative
techniques.\cite{wilson,le_guillou_zinn-justin,3DIsing2,free_params2RG,free_params2RG2}

Of course, the omission of nonlocal contributions even at large
$t_0$/small $k_0$ values will still introduce some errors into the
results. But it should be remembered that the cluster part of the
calculations will always be of restricted accuracy because the complexity
of the calculations grows exponentially with the cluster size, so the
size will be bounded by some finite value of the order of those used in
the calculations of Ref.\ \onlinecite{tan_topologically_2011}. So making
the RG part much more accurate does not have much sense.
\subsection{\label{observables}Measurable quantities} 
With Eq.\ (\ref{the_eq}) the field-theoretic problem with infinite number
of variables reduced to a Cauchy's problem for the second-order evolution
equation for a function of two variables.  From Eqs.\ (\ref{U0}), (\ref{U(k)}),
and (\ref{U2u}) it is easy to derive the initial condition
\begin{equation}
u(x,t_0)=\bar{G}_{ii}^{-1}x^2/2-\ln\cosh(\bar{G}_{ii}^{-1}x)+C
\label{u(x,t_0)}
\end{equation}(we remind that in the RG part we set $m=0$ both above and below
the critical temperature).  The only peculiarity of our Cauchy problem is that
it should be solved in a self-consistent manner to satisfy 
the condition Eq.\ (\ref{d2rdphi2}):
\begin{equation}
	u_{xx}(x=h/r,t=1/r)=0.
	\label{u_xx=0}
\end{equation}

To calculate physical quantities from in the SSA+LPA we first need to
find the per-site Gibbs potential 
\begin{equation} 
	g=-\ln Z[h]/N  
	\label{g} 
\end{equation}
(we remind that $k_BT=1$ in our units).  In the LPA the field $h_i$, that
in principle could be used to calculate via the differentiation of $Z[h]$
all correlation functions, reduces to a single number---the value of the
zero-momentum component, i.\ e., to the homogeneous external magnetic
field $h$ (we will denote it by the same letter because from now on
it will have only this meaning).  So only homogeneous responses can be
found within LPA.  It is important to point out, however, that in the
self-consistent approach the pair correlation function can also be
calculated through Eqs.\ (\ref{G}) and (\ref{u_xx=0}) because the first
equation was derived before the local approximation was made.

With the use of Eq.\ (\ref{ZR}) in the SSA+LPA approximation one gets
\begin{equation} 
	g(h) = -\frac{1}{2N}\ln\det G-\frac{h^2}{2r}+u(x=\frac{h}{r},t=\frac{1}{r}),
\label{g_LPA} 
\end{equation} 
To find solutions with spontaneously broken symmetry, however, we need
the Helmholtz free energy that can be obtained from the Gibbs potential
in a standard way via the Legendre transform:\cite{zia,vasiliev1998}
\begin{equation} f(m)=g-hg_h=g+hm,
\label{helmholtz} 
\end{equation} 
where the dependence of $h$ on magnetization $m$ is found from the definition
\begin{equation} m = -g_h = x-u_x/r.
\label{m} 
\end{equation} 
The magnetic susceptibility is
\begin{equation} \chi=m_h=-g_{hh}=r^{-1}-u_{xx}/r^2=r^{-1},
\label{chi} 
\end{equation} 
where use has been made of the self-consistency condition Eq.\ (\ref{u_xx=0}).  We note
that in our approximation Eq.\ (\ref{chi}) agrees with another definition
\begin{equation} \chi=G({\bf k = 0})=1/r
	\label{chi1} \end{equation} 
so that at this level of approximation the approach is consistent.

The critical amplitudes $C^{\pm}$ describe the asymptotic
behavior of the susceptibility in the vicinity of the critical point
$K_c$ as\cite{liu_fisher89,Wagner1994,RG2002review}
\begin{equation}
	\chi\simeq C^{\pm}\tau^{-\gamma}, 
	\label{defCpm}
\end{equation}where 
\begin{equation}
	\tau = |K_c/K-1|
	\label{tau}
\end{equation}
$\gamma$ is the critical index, and $\pm$ denotes whether the amplitude
is being measured or calculated above (+) or below (-) the critical
temperature.  Now because $r$ in Eqs.\ (\ref{chi}) and (\ref{chi1}) can be
found from the solution of the SSA+LPA equations, the amplitudes in Eq.\
(\ref{defCpm}) can be calculated.  Similarly, from the Fourier transform
of Eq.\ (\ref{G}) the correlation length can be found as
\begin{equation}
\xi = \sqrt{K/r}\simeq f^{\pm}\tau^{-\nu}, \label{f+} \end{equation} 
where $\nu$ is another critical index.  From this relation the behavior
of $r$ in the critical region can be derived as
\begin{equation} r = a\tau^{2\nu}, \label{r}
\end{equation} 
where $a$ is a constant.  Comparing this with Eq.\ (\ref{defCpm}) one can
find $a=1/C^{\pm}$ and the relation between the exponents is
\begin{equation}
\gamma = 2\nu.
\label{gamma=2nu}
\end{equation}
This relation imply $\eta=0$\cite{RG2002review} and is a consequence of
the LPA.\cite{1984}

Finally, the amplitude $B$ and the critical index $\beta$ are defined from
the behavior of the spontaneous magnetization near the critical point as
\begin{equation}
	m_0\sim B\tau^{\beta}.
	\label{m_0}
\end{equation} 
These five amplitudes ($C^\pm, f^\pm$ and $B$) together with the critical
temperature will be the non-universal quantities that we will calculate
within the C+RG approach in the present paper.

Eq.\ (\ref{the_eq}) is easily solvable numerically in the symmetric
phase.  So if we were interested only in critical amplitudes, we could
solve Eqs.\ (\ref{the_eq}) and (\ref{u_xx=0}) in this phase
to find $C^+$ and $f^+$ and then utilized known values of the 
universal ratios to calculate $C^-$, $f^-$ and even
\begin{equation} 
	B=\sqrt{Q_cC^+/f^+}, \label{B}
\end{equation} 
where $Q_c$ is a known universal quantity.\cite{RG2002review} However,
because our aim is to develop a method that would describe thermodynamics
also beyond the critical region, we would like to solve the C+RG
equations below the critical temperature as well.  Unfortunately, Eq.\
(\ref{the_eq}) is not easily solvable in the ordered phase.  Therefore,
numerical calculations were performed with transformed equation that
could be solved in both phases as explained below.
\section{\label{numerics}Numerical solution of the LPA equation} 
Numerical solution of a Cauchy problem for equations similar to Eq.\
(\ref{the_eq}) is usually reduced to solution of a set of ordinary
differential equations via the method of lines. \cite{Hamdi:2007} In
the disordered phase Eq.\ (\ref{the_eq}) can be quite efficiently solved
with the use of this method but in ordered phase it becomes practically
intractable, at least at the double precision arithmetics that was used in
the calculations.  Apparently, the reason for this lies in the RG
nature of the equation. As such, Eq.\ (\ref{the_eq})
describes the renormalization of all Hamiltonians of the LGW type,
in particular, the quadratic Hamiltonian corresponding to the exactly
solvable Gaussian model.\cite{wilson}  In our formalism this shows itself
in the existence of an exact solution of Eq.\ (\ref{the_eq}) of the form
\begin{equation} u^{(G)}(x,t)=u_2x^2+u_1x+u_0. \label{u_G}
\end{equation} 
As is known, the Gaussian model is pathological in that at negative $u_2$ 
it is not bounded from below and so does not possess an ordered state.
So if one attempts integration of Eq.\ (\ref{the_eq}) with the initial
$u^{(G)}(x,t_0)$ having negative value of $u_2(t_0)$, than the solution 
that can be obtained analytically in this case will have the form
\begin{equation} u_2(t) = \frac{1}{2t+c_0},
\label{u2_G} 
\end{equation} 
where $c_0$ is a negative constant such that $1/(2t_0+c_0)=u_2(t_0)<0$.
As we see, at some later time $t=-c_0/2$ Eq.\ (\ref{u2_G}) develops a
non-integrable singularity $u_2\to\-\infty$ that cannot be treated within
conventional finite-difference schemes.  This is what was qualitatively
observed in calculations.  Obviously that negative $u_2(t=t_0)$ are
forbidden in the Gaussian model because the probability distribution
became non-normalizable.  But negative $u_2$ is exactly what drives
the ordering in the interacting LGW model.  The Hamiltonian is bounded
from below due to quartic and higher terms so the initial condition with
$u_2(t=t_0)<0$ is what happens in reality.  But in numerical calculations
due to computational errors and the finite range of variation of $x$
the trajectory leading to unphysical solution is apparently very close
to the physical trajectory so that their separation is numerically
proved to be impossible.

In Ref.\ \onlinecite{local_potential} time-independent analogue of Eq.\
(\ref{the_eq}) in the scaling limit ($r=0, t\to\infty$, $s$ corresponding
to the square root density of states) in scaling variables\cite{1984} was
studied at the critical point and also was found to be very difficult to
deal with numerically.  It was found, however, that a change of variables
can significantly alleviate the difficulties.  It turned out that with
necessary modifications this change of variables can be helpful also in
our case of time-dependent Eq.\ (\ref{the_eq}) in the non-scaling regime.
Specifically, by introducing new independent variable $y$ and new function
$v(y,t)$ as
\begin{subequations}
	\label{eq:all}
\begin{eqnarray}
	&&y(x,t)=x - (t-c)u_x(x,t)\\ \label{eq:a}
	&&v(y,t)=u(x,t)-{(t-c)}u_x^2/2, \label{eq:b}
\end{eqnarray} \end{subequations} ($c$ is an arbitrary constant)  
Eq.\ (\ref{the_eq}) can be cast in the form (see Appendix \ref{y-v})
\begin{equation}
	v_t = \frac{s(t,r)v_{yy}}{2[1+(t-c)v_{yy}]}  \label{the_eq2}
\end{equation} 
with the self-consistency condition
\begin{equation}
v_{yy}=0
\label{v_yy=0}
\end{equation} (see Eq.\ (\ref{u_xx-v_yy})).
From Eq.\ (\ref{the_eq2}) it is easy to see that the quadratic model of the kind
of Eq.\ (\ref{u_G}) but expressed in terms of the new variables does not
exhibit any singularities because $\partial v_2(t)\partial t=0$ so the
initial value of $v_2$ remains the same irrespective of its sign.
This solves the problem of the singular second derivative and allows
for numerical solution in both symmetric and ordered phases.

Equations similar to Eq.\ (\ref{the_eq2}) were
studied previously in LPA-RG approaches (see Refs.\
\onlinecite{local_potential,caillol_non-perturbative_2012} and references
therein) with some physical meaning being ascribed to the quantities
similar to our $v$ and $y$.  In our approach, however, such a meaning
is not easily discernible so we will consider these entities as purely
auxiliary variables while physical quantities will be calculated with
the use of thermodynamic potentials introduced in previous section.

As is easily seen, the arbitrary constant $c$ in Eqs.\ (\ref{eq:all})
is convenient to chose to be equal to $t_0$.  In this case at $t=t_0$
$y=x$ and $v=u$, so the initial function $v$ is the same as in Eq.\
(\ref{u(x,t_0)}) with only $x$ replaced with $y$.  Thus, denoting
the time span from the start to the end of the integration as
\begin{equation}
\Delta t = (1/r-t_0) 
\label{delta_t}
\end{equation}
with the use of Eqs.\ (\ref{g_LPA}), (\ref{m}), (\ref{chi}), (\ref{eq:all}),
and Appendix \ref{y-v} we can express all physical quantities in terms of
the auxiliary variables as
\begin{subequations}
   \label{f_fm_fmm} \begin{eqnarray}
      && x = h/r = y+\Delta t v_y \\ 
      \label{h}
      && u = v+\Delta t
      v_y^2/2 \\ 
      && m = y - t_0v_y\\ 
      && f = g - xg_x = u-xu_x+rx^2/2\\ 
\label{f}
      && \chi^{-1} = f_{mm} = r(1+\Delta
      t v_{yy})/(1-t_0v_{yy})=r,
\label{sc}
   \end{eqnarray}
\end{subequations}
where in the last equality use has been made of the self-consistency
condition Eq.\ (\ref{v_yy=0}).
\subsection{\label{SCIsing}SC Ising model in zero magnetic field}
\begin{figure}
\begin{center}
\includegraphics[viewport = 80 10 300 210, scale=0.75]{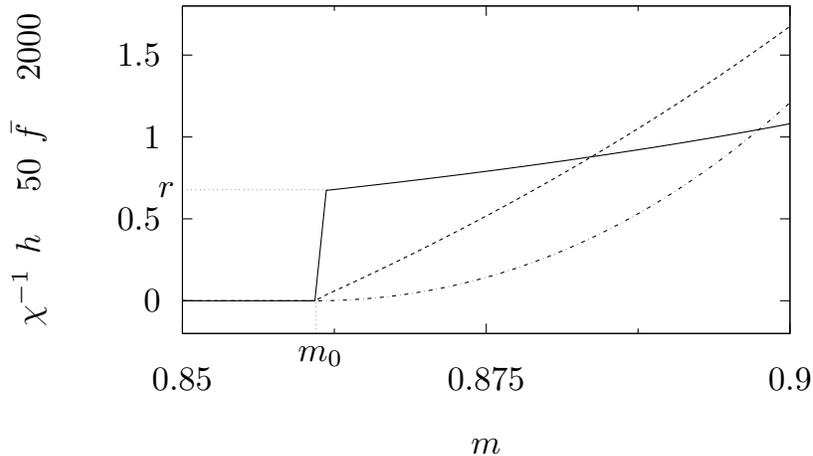}
\end{center}
\caption{\label{fig4}Dependence of some thermodynamic quantities on 
magnetization in the vicinity of the first-order phase transition from the
positive to the negative equilibrium magnetization values. For clarity,
only small part of the graphs near $m_0>0$ is shown. Solid line: the
inverse susceptibility $\chi^{-1}$ Eq.\ (\ref{sc}); dashed line: external
field $h$ Eq.\ (\ref{h}); dashed-dotted line: $f(m)-f(0)$, where $f$ is the 
free energy density Eq.\ (\ref{f}).}
\end{figure}

Because our major interest in the present paper is in the second-order
phase transitions, we will restrict the solution of the Ising model
to the $h=0$ case.  In the symmetric phase $v(y,t)$ is a smooth
symmetric function and its numerical solution is rather unproblematic.
The difficult part of the problem was the self-consistent solution of
Eqs.\ (\ref{the_eq2}) and (\ref{v_yy=0}) in the ordered phase where the
solution exhibited non-smooth behavior (see Fig.\ \ref{fig4}).  Both the
differential equation {\em per se} and the self-consistency requirement
posed problems. In Ref.\ \onlinecite{caillol_non-perturbative_2012}
the author resorted to quadruple-precision computations to reliably
solve equations of this kind with the use of the method of lines.
Unfortunately, in the present study such facilities were unavailable,
so conventional double-precision arithmetic and the LSODE routine at the
highest possible accuracy (the absolute and the relative tolerances set
equal to the machine epsilon\cite{lsode}) were used in the calculations.
Fortunately, in Ref.\ \onlinecite{caillol_non-perturbative_2012}
it was established that the discretization step in magnetization
equal to $2\cdot10^{-3}$ is already sufficient to obtain the correct
solution. This kind of accuracy was accessible to the software used
so solutions obtained was in qualitative agreement with previous
studies.\cite{maxwell_construction,caillol_non-perturbative_2012}

The solution converged to reproducible results with the use of the
following parameters.  Eq.\ (\ref{the_eq2}) was divided into 2200
ordinary evolution equations: 2000 covering the range [0,2] along the
coordinate $y$ with the step $10^{-3}$ and 200 equations in the range
(2,5] with the step 0.015.  The difference in the step size was due to
different smoothness of $v(y,t)$ in different regions.

\begin{figure}
\begin{center}
\includegraphics[viewport = 150 20 400 300, scale=0.5]{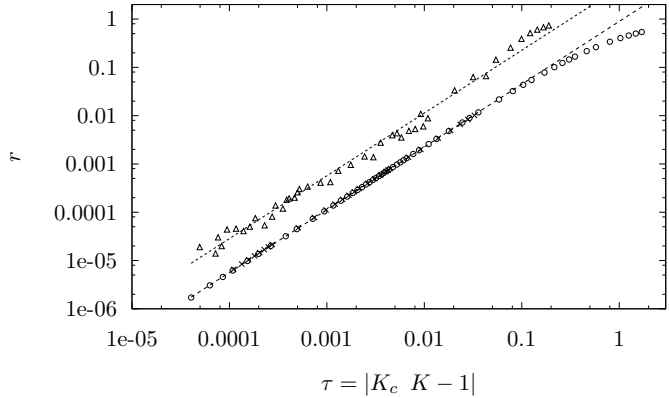}
\end{center}
\caption{\label{fig5}Behavior of $r$ near the critical point in
the symmetric (circles) and ordered (triangles) phases for $t_0=0.65$;
crosses---$r$ in the symmetric phase for $t_0=2$. Straight lines are
fits to Eq.\ (\ref{r}) for the data in the region $\tau<10^{-3}$.}
\end{figure}

The most difficult region of the solution was the jump in the inverse
susceptibility that was observed in the data (see Fig.\ \ref{fig4}).
In contrast to Ref.\ \onlinecite{maxwell_construction}, the naive
calculation of $C^+/C^-$ ratio on the basis of our data from Table
\ref{the_table} gives value 5 in excellent agreement with the best
estimates of this universal ratio.\cite{RG2002review} This result,
however, has to be taken with caution because of the large error in the
calculated $C^-$ (see Table \ref{the_table}).  The reason for this was
the self-consistency requirement Eqs.\ (\ref{v_yy=0}) and (\ref{sc})
according to which $r$ has to have the value equal to the jump in the
inverse susceptibility.  Geometrically this means that the value of
the curve in Fig.\ \ref{fig4} at the upper side of the step should
be equal to $r$.  But in fitting $r$ to fulfill this requirement one
meets with the problem that the position $m_0$ of the jump could change
only in finite steps $\sim2\cdot10^{-3}$ (see above).  Because of this,
it was impossible to make $r(\tau)$ curve in Fig.\ \ref{fig5} to be
sufficiently smooth to ensure an accurate definition of $C^-$.
\begin{table}
        \caption{Comparison of the C+RG calculations of several non-universal 
		quantities for the SC Ising model with the series expansions and 
		the MC simulations data of Refs.\ \onlinecite{liu_fisher89} and 
		\onlinecite{Wagner1994}, respectively}
 \label{the_table}
        \begin{tabular}{c|c|c|c|c|c|c|}
            \cline{2-7}
	    & $K^{-1}_c$& $C^+$ & $f^+$ & $C^-$ & $f^-$ & $B$ \\
            \hline
\multicolumn{1}{|c|}{SSA+LPA} 
&4.44 &1.1 & 0.50&0.22$^{\rm a}$&0.22$^{\rm a}$&1.6\\
            \hline
	    \multicolumn{1}{|c|}{Refs.\ \onlinecite{liu_fisher89}$^{\rm b}$ 
	    and \onlinecite{Wagner1994}$^{\rm b,c}$} 
& 4.51&1.1 &  0.49--0.50 &0.21--0.22&0.24--0.25&1.6--1.7\\
            \hline
        \end{tabular}\\
	\begin{flushleft}
	$^{\rm a}$The values are connected with the calculations of $r$ 
		in the ordered phase, so their accuracy is estimated to 
		be $\sim20\%$ (see the text for details)\\
	$^{\rm b}$To facilitate comparison, the data were rounded to the 
	percieved accuracy of the SSA+LPA\\
	$^{\rm c}$ Only the data recommended by the authors were used
	\end{flushleft}
    \end{table}
In this connection a more general question arises about the very {\em
existence} of the jump. It was found that its presence or absence
depends strongly on the renormalization scheme used and on the system's
dimensionality.\cite{maxwell_construction,caillol_non-perturbative_2012}
The calculations presented in Fig.\ \ref{fig4} were performed at
the largest value of $r=0.68$ in our calculations in order to have
the smallest integration time span $\Delta t = 1/r-t_0\approx 0.82$.
But even at such a short integration interval the jump was already
sufficiently abrupt to fall within only one $\Delta m$ step.  Besides,
further properties of the numerical solution obtained were found
to be in qualitative agreement with the known properties of the exact
solution.\cite{vasiliev1998,maxwell_construction,caillol_non-perturbative_2012}
Namely, inside the mixed-phase region where magnetization
varied from $-m_0$ to $m_0$ the free energy $|f|$ varied
on less than $3.5\cdot10^{-7}$, $|h|<7.5\cdot10^{-7}$, and
$|\chi^{-1}|<5.3\cdot10^{-5}$. The inverse susceptibility variation was
quite random and largely due to the computational errors.  In the exact
theory all these values on the horizontal line in Fig.\ \ref{fig4}
should be equal to zero.

Of course, in numerical calculations it is impossible to prove, for
example, that there is the abrupt jump in the susceptibility and not
only very steep region on the curve.  But there seems to be no need for
this because the calculations anyway are only approximate.  So even if
there is only very steep curve, it will only be necessary to develop
some procedure similar to Maxwell's construct to calculate the most
probable position and the gap size of the jump that is known to take
place in the system in reality.  Presumably, more rigorous possibility
would be to adjust the renormalization scheme to have a real jump. This,
however, will not make the calculations more precise and it should be
remembered that exact results are renormalization-scheme independent.
More thorough investigation of this question is needed to take finite
decisions on the subject.

\begin{figure}
\begin{center}
\includegraphics[viewport = 150 20 400 200, scale=0.75]{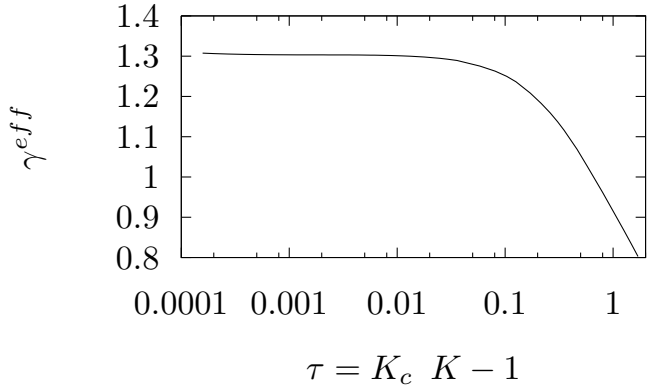}
\end{center}
\caption{\label{fig6}Effective critical exponent $\gamma$ fitted
to Eq.\ (\ref{chi1}) for different fitting intervals $\tau$.  No region
of mean-field behavior with $\gamma=1$ could be discerned.}
\end{figure}

The results of the calculations are presented in Figs.\
\ref{fig1}, \ref{fig4}, \ref{fig5}, \ref{fig6}
and in Table \ref{the_table}.  The critical quantities in the Table were
derived from the fit to Eqs.\ (\ref{defCpm}), (\ref{f+}), (\ref{r}), and
(\ref{m_0}).  In all fits except one, both the amplitudes and the critical
indexes were fitted to the data.  The exception was made for $r$ in the
ordered phase where the scatter in the data (see Fig.\ \ref{fig5})
for the reasons explained above made the two-parameter fit unstable.
Therefore, only the amplitude was fitted in Eq.\ (\ref{r}) for this case
while the critical index was given the known LPA value.  In all other
cases in sufficiently small region near the critical point (see Fig.\
\ref{fig6}) the indexes coincided to a good accuracy with the
LPA values.  The latter were derived from the known $\eta^{LPA}=0$
and $\nu^{LPA}\simeq0.65$\cite{1984,local_potential} and the scaling
relations as $\gamma^{LPA}=2\nu^{LPA}$ and $\beta^{LPA}=\nu^{LPA}/2$.

To assess the sensitivity of the results to the choice of large $t_0$,
the SSA+LPA equations were solved in the disordered phase for $t_0=2$.
As can be seen from Fig.\ \ref{fig5}, the dependence $r(\tau)$ (hence,
$\chi(\tau)$---see Eq.\ (\ref{chi})) remained practically unchanged.
The agreement of the calculated critical temperature $K^{-1}_c|_{t_0=2}=4.26$
with the exact MC value worsened but only on $\sim4\%$ with respect
to the $t_0=0.65$ case which is not too drastic a change taking into
account that the cutoff $t_0$ changed more than three times.

In Fig.\ \ref{fig1} the boundary between ordered and disordered phases was
drawn according to the SSA solution by starting from low temperatures and
according to the LPA solution when starting from the critical region.
The curves met at the point $1/K\approx3.7$.  Despite a small cusp,
the combined curve agrees well with the exact MC simulations of Ref.\
\onlinecite{talapov_M(t)} and with the exact low-temperature asymptotic
behavior.\cite{lowT_sc} As is seen, the critical behavior smoothly
passes into the low-temperature regime and no mean-field square-root
like behavior can be discerned.  Similar absence of the mean-field regime
took place in our calculations above the critical temperature (see Fig.\
\ref{fig6}).  It would be interesting to check this predictions with
MC simulations or series expansions because it would allow to check
the ability of our approach to describe the transient region between
the critical point and the far from criticality region also in the
symmetric phase.

The largest discrepancy in the SSA+LPA calculations was observed in the
calculation of $r$ at temperature $K_1^{-1}$ (see Fig.\ \ref{fig1}).
The SSA and LPA solutions differed more than tree times:
\begin{equation}
	r_1^{SSA}\approx 2.36\;\mbox{ and }r_1^{LPA}\approx 0.68.
	\label{r1}
\end{equation} 
It is not clear which part of the solution to blame because at their
respective temperature regions both solutions agree well with the best
known solutions.  Further work is needed to clarify this problem.

In connection with Table \ref{the_table} it should be noted that
several other amplitude values can be calculated from the data with
the use of the universal amplitude ratios known from the universality
studies. \cite{RG2002review}  This may be useful in practice when only
critical amplitudes are of interest but direct derivation of some of
them from the C+RG solutions is difficult.

\section{\label{conclusion}Conclusion}
The main result of the present paper is the development of the technique
that allows for quantitative description of the second order phase
transitions from known microscopic interactions of arbitrary strength.
The layer-cake renormalization scheme allowed us to calculate the
critical temperature with better than 2\% accuracy and several critical
amplitudes in good agreement with MC simulations and series expansions.
The description of the critical regime in the LPA, however, was not
perfect.\cite{1984} The improvement of the method should apparently lie in
going beyond the LPA and accounting for non-local/quasimomentum-dependent
contributions.  This should also improve the critical indices that are
known to be not very good because, for example, $\eta^{LPA}=0$ due to the
locality \cite{1984,local_potential,caillol_non-perturbative_2012}. But
more important would be the possibility to account for the non-pair
cluster interactions that are inherently nonlocal.

The main domain of application of the technique developed
in the present paper is envisaged in the theories of
phase transitions based on realistic interactions either
calculated {\em ab initio} or extracted from experimental
data.\cite{ECI,ducastelle,gautier,c_INdep_CE2,Zunger1994,EPIs,PhysRevLett.92.255702}
Such interactions usually do not possess high symmetry and simplicity of
theoretical models.  Therefore, C+RG was designed to be able to account
in a brute force manner for as many various interactions as possible.
The cluster methods in principle are well suited to this goal but,
unfortunately, they are restricted to rather large quasimomentum cutoffs.
For example, the $4^d$ hypercubic cluster we used in the discussion
is already highly unrealistic in 3D ($\sim2^{64}$ terms would need be
accounted in the cluster equations) though the quasimomentum cut-off
provided by it will be only one fourth of the initial (maximum) cut-off.
Therefore, some non-negligible nonlocalities may remain and it would
be desirable to take them into account at the RG stage of the method.
It should be born in mind, however, that due to smaller quasimomenta and
smaller renormalized coupling constants, the nonlocality may be strongly
suppressed at the RG stage if sufficiently large clusters are used at the
cluster stage.  So apart from the incorrect critical indexes the technique
can be quite accurate for large clusters even in its current form.

As to the critical behavior {\em per se}, it seems to be unrealistic to
demand from a theory that aims to take into account as many realistic
microscopic interactions (cluster, long-range, etc.) as possible,
to be equally successful at the other extreme of the essentially
macroscopic critical behavior.  Presently to describe the critical region
oversimplified models are being treated within the most sophisticated and
rigorous series expansion techniques, special field-theoretic models are
being designed to suppress undesirable contributions that obscure the
behavior of interest, and yet disagreements and discrepancies between
different studies continue to exist.\cite{RG2002review}  Therefore,
instead of trying to solve some improved LPA equations in the critical
region to obtain {\em a priori} inferior results, it seems more
reasonable to integrate the equations up to some preasymptotic boundary
to find the few parameters needed in the quantitative description of the
critical behavior in the framework of some sophisticated field-theoretic
approach.\cite{free_params2RG,free_params2RG2}

Further work is necessary to develop practically efficient combination of
cluster and RG methods to accurately describe thermodynamics of strongly
coupled field-theoretic statistical models.  The major conclusion that
can be drawn from the results of the present study is that this task is
quite feasible.
\acknowledgments
I am grateful to Hugues Dreyss\'e for encouragement.
\appendix
\section{\label{Layer_Cake}Layer cake renormalization}
The layer cake representation \cite{layer_cake} is, essentially, based on the identity
\begin{equation}
	G = \int_0^{\infty}\theta(G-t)dt.
	\label{G-theta}
\end{equation}
valid for positive $G$. To our purposes it is more suitable in differential form
\begin{equation}
	dG=\theta(G-t)dt.
	\label{dGdt}
\end{equation}
Theta-function in its turn can be represented as
\begin{equation}
	\theta(G-t)=\int_t^{\infty}\delta(G-t^\prime)dt^\prime.	
	\label{theta-delta}
\end{equation}Assuming now that $G$ is a function of $\epsilon$ 
\begin{equation}
	G = \frac{1}{\epsilon+r}
	\label{G2}
\end{equation}
(cf.\ Eq.\ (\ref{G})) and using the identity
\begin{equation}
	\delta\left(f(x)\right)=\frac{1}{|f^\prime(x_0)|}\delta(x-x_0)
	\label{delta-delta}
\end{equation}Eq.\ (\ref{theta-delta}) can be transformed as
\begin{equation}
	\theta(G-t)=\int_t^\infty\frac{dt^\prime}{{t^\prime}^2}\delta\left(\epsilon-\frac{1}{t^\prime}+r\right)
	=\int_0^{t^{-1}-r}dE\delta(E-\epsilon),
	\label{finalG-t}
\end{equation}where $\epsilon$ and $r$ are assumed to be non-negative.

Because in the LPA the vertex functions are ${\bf k}$-independent, the
summation over quasimomentum in Eq.\ (\ref{linear_term}) can be performed
as
\begin{equation}
	\frac{1}{N}\sum_{\bf k}\theta\left[ G({\bf k},t)-t\right]
	=\int_0^{t^{-1}-r}dE\frac{1}{N}\sum_{\bf k}\delta[E-\epsilon({\bf k})]
        =\int_0^{t^{-1}-r}dED(E),
	\label{sum2D(E)}
\end{equation}
where $D(E)$ is the density of states in the quasiparticle band with
the energy dispersion $\epsilon({\bf k})$ (see Fig.\ \ref{fig3} for
illustration of Eq.\ (\ref{sum2D(E)}) for the case of the SC Ising
model).  \section{\label{y-v}Change of variables} Differentiating Eqs.\
(\ref{eq:all}) with respect to $x$ one gets:
\begin{subequations}
	\label{eq:all_prime}
\begin{eqnarray}
	&&y_x=1 - (t-c)u_{xx}\\	
	\label{eq:a_prime}
	&&v_yy_x=u_x[1-{(t-c)}u_{xx}]
	\label{eq:b_prime}
\end{eqnarray}
\end{subequations}
which gives, in particular,
\begin{equation}
v_y=u_x.
\label{v_y=u_x}
\end{equation}Now, taking the first derivative of Eq.\ (\ref{v_y=u_x}) with respect to $x$ and using Eq.\ 
(\ref{eq:a_prime}) one arrives at
\begin{equation}
u_{xx}=\frac{v_{yy}}{1+(t-c)v_{yy}}.
\label{u_xx-v_yy}
\end{equation}Next, by differentiating Eq.\ (\ref{eq:all}) with respect to $t$ one obtains with
the use of Eq.\ (\ref{v_y=u_x})
\begin{equation}
u_t=v_t-u_x^2/2.
\label{u_t-v_t}
\end{equation}
By substituting Eqs.\ (\ref{u_xx-v_yy}) and (\ref{u_t-v_t}) in Eq.\ (\ref{the_eq})
one arrives at Eq.\ (\ref{the_eq2}).
\bibliographystyle{apsrev} 

\end{document}